\documentclass[aps,preprint,showpacs]{revtex4}
\usepackage{graphicx}
\usepackage{latexsym} 
\usepackage{amsmath,amscd}
\usepackage{times}
\usepackage{epstopdf}

\begin{document}

\title{\bf\noindent Capture of particles undergoing discrete random walks}
\author{Robert M. Ziff,$^{1,2}$ Satya N. Majumdar,$^2$ and  Alain Comtet$^{2}$}
\affiliation{
$^1$ Michigan Center for
Theoretical Physics and Department of Chemical
Engineering, University of Michigan, Ann Arbor, MI 48109-2136 USA \ 
$^2$Laboratoire de Physique Th\'eorique et 
Mod\`eles Statistiques (UMR 8626 du CNRS), 
Universit\'e Paris-Sud, B\^at.\ 100, 91405 Orsay Cedex, France 
}
\pacs{02.50.-r, 02.50.Sk, 02.10.Yn, 24.60.-k, 21.10.Ft}

\begin{abstract}
It is shown that particles undergoing discrete-time jumps in $3$D, starting at a distance $r_0$ from the
center of an adsorbing
sphere of radius $R$, are captured with probability
$(R- c \sigma)/r_0$ for $r_0 \gg  R$, where $ c$ is related to the 
Fourier transform of the scaled jump distribution and $\sigma$ is the distribution's root-mean square jump length.
For particles starting on the surface of the sphere, the asymptotic survival probability
is non-zero (in contrast to the case of Brownian diffusion) and has a universal
behavior  $\sigma / (R \sqrt{6})$ depending only upon $\sigma/R$.
These results have applications to computer simulations of reaction and aggregation.

\end{abstract}
\maketitle

\section{Introduction}

A celebrated result in probability theory states that in three or higher dimensional space,
particles undergoing diffusion or Brownian motion have a chance to escape to infinity
rather than be captured to a finite adsorbing surface.   This was first demonstrated in the
case of random walks (discrete versions of Brownian motion) on
a $3$D cubic lattice, on which  a walker 
starting at a site adjacent to the adsorbing origin escapes with probability
$I^{-1}\approx 0.659463$ or is captured with probability $1-I^{-1} \approx 0.340537$,
where \cite{Polya21,Watson39,GlasserZucker77}
\begin{eqnarray}
I &=& \frac{3}{(2 \pi)^3} \int_{-\pi}^\pi  \int_{-\pi}^\pi  
\int_{-\pi}^\pi \frac{dx\,dy\,dz}{3 - \cos x - \cos y
- \cos z} \nonumber \\
&=& \frac{\sqrt{6}}{32 \pi^3} \Gamma\left(\textstyle \frac 1 {24}\right)  
\Gamma\left(\textstyle \frac 5 {24}\right) 
 \Gamma\left(\textstyle \frac 7 {24}\right)  \Gamma\left(\textstyle \frac {11} {24}\right) \ .
 \label{cubic}
\end{eqnarray}  
In fewer than three dimensions, walks return to the origin an infinite number of times and
are thus recurrent; while in three and higher dimensions they escape and are
transient.   The transient nature of random walks in $3$D
is the basis for long-range diffusive transport and is essential for many
important physical phenomena involving particles that must escape
a locally trapping region.

The same phenomenon in continuum theory is illustrated by the problem of adsorption of
diffusing point particles on a spherical boundary of radius $R$.  Say the particle starts initially at 
a distance $ |{\bf r}_0|\ge R$ from the center of a sphere. The survival probability
$S({\bf r_0},t)$ up to time $t$ satisfies the backward diffusion equation
$\partial S({\bf r}_0,t) / \partial t = D \nabla^2 S({\bf r}_0,t)$ with
the initial condition $S({\bf r}_0,0) = 1$ for all $|{\bf r}_0| > R$ and the boundary conditions
$S(|{\bf r}_0| = R,t) = 0$ and $S(|{\bf r}_0| \to \infty, t) = 1$ for all $t$. 
Exploiting radial symmetry, the solution to this 
equation is easily found at all times $t$
\begin{equation}
S(r_0, t) =  1 - \frac R {r_0} {\rm erfc} \frac{r_0 - R}{\sqrt{4 D t}}. 
\label{survival}
\end{equation}
Taking the limit $t \to \infty$, one finds that the probability
to ultimately escape 
is simply
\begin{equation}
S(r_0,\infty) = 1 - R/r_0 \ ,
\label{escape}
\end{equation}
a well-known result in probability theory.  
The capture probability is  $P(r_0, \infty) = 1 - S(r_0,\infty) = R/r_0$.

In this paper, we ask how having a finite step-length in the diffusing
particle's motion  and discrete time-steps
affects the continuum results in Eqs.\ (\ref{survival},\ref{escape}).
In particular, we consider a random walker in $3$D
continuum where, in each step of equal time interval $\tau$, the particle jumps isotropically a 
distance $r$ drawn from a distribution $4 \pi r^2 W(r)$, bounded above by $2R$, but 
otherwise arbitrary. If the particle jumps inside the sphere it is captured.
Let $S_n(r_0)$ be the 
probability that the particle, starting at $r_0\ge R$,
survives up to step $n$. The net capture probability is then $P_{\infty}(r_0)=1-S_{\infty}(r_0)$. \
The question of the net capture probability in this case is
also important to many computer simulation problems, where particle diffusion is
carried out by such a discrete-time process, typically through a Pearson flight
in which the particle at each time step jumps a discrete distance $\ell$
which may be of the order of magnitude of $R$ or even larger.  The
capture probability can be used to find chemical or catalytic reaction rates, and is also 
relevant to non-equilibrium growth studies, notably in diffusion-limited aggregation
\cite{WittenSander81} and related aggregation models.  There has been
little (or no) discussion on how the simulation's jump length affects the
ultimate results of these simulations.

Another situation where a finite jump length is relevant is when particles are
diffusing in a gas and the mean free path becomes comparable to the diameter
of the adsorbing sphere.  Then the motion of the diffusing particle is not an
infinitesimal Wiener process described by the diffusion equation, but instead is
described by the Ornstein-Uhlenbeck theory of Brownian motion with inertial drift, 
integrated to position space \cite{UhlenbeckOrnstein30}.
Only after particles travel
a certain distance do they ``forget" their initial direction and continue in a random direction.  
An approximation to this process is given by discrete random walk.

\begin{figure}
\includegraphics[width=0.3\hsize ]{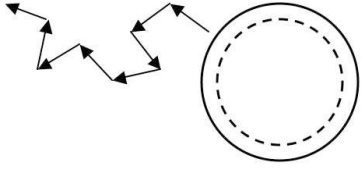}
\caption{Example of a walk that starts near the surface of the sphere
of radius $R$ and eventually
escapes.  The dashed circle represents the effective capture radius
$R' = R -  c \sigma$
for particles that start far from the sphere.}
\label{walkfig}
\end{figure}

For the discrete random walk problem, we find that the survival probability $S_n(r_0)$ 
becomes substantially modified from its continuous-time counterpart $S(r_0,t)$.
We show that the ultimate survival probability $S_{\infty}(r_0)$ has
the following asymptotic behavior
\begin{eqnarray}
S_{\infty}(r_0) & \approx & 1- \frac{R-c\sigma}{r_0}\quad {\rm for} \,\, r_0\gg R \nonumber \\
&\to & \frac{\sigma}{R \sqrt{6}}\quad\quad\quad\quad {\rm as}\, \,\, r_0\to R
\label{univ}
\end{eqnarray}
where $\sigma^2= 4\pi \int W(r) r^4 dr$
is the mean-square jump distribution and $c$ is a constant (computed explicitly below) that 
depends upon 
the details of $W(r)$. 
Thus if the particle starts very far from the surface of the sphere, the final
survival probability has a similar expression as in the continuous-time case in
Eq.\ (\ref{escape}), except that the effective radius of the sphere $R'=R-c\sigma$ 
shrinks
by a finite `extrapolation' length $c\sigma$ (see Fig. \ref{walkfig}). On the other hand, if the
particle starts right outside the surface of the sphere we find
a rather surprising result: in sharp contrast to the continuous-time case,
the survival probability is nonzero and moreover has a universal value
$1/\sqrt{6}$ in units of the dimensionless ratio $\sigma/R$, and
is completely independent of the other details of the jump distribution (see Fig. \ref{fig2}). 


\section{Derivation of results}

To derive these results our strategy is to first map the capture problem
to the so called `flux' problem to a sphere and then use the known results
of the latter problem. This mapping works both
for continuous-time Brownian motions as well as for discrete-time random walkers.
For simplicity we discuss first the mapping for continuous-time Brownian motions.
The continuous-time flux problem, first studied by   
Smoluchowski \cite{Smoluchowski16,Chandrasekhar43}, is defined as follows.
Consider an infinite number of noninteracting Brownian particles initially
distributed with uniform density $\rho_0$ outside a sphere of radius $R$ in
$3$D. Each particle subsequently performs Brownian motion and, if it reaches
the sphere, it is absorbed. If $\rho(r,t)$ denotes the density profile
at time $t$, the instantaneous flux to the sphere is $\Phi(t)=4\pi R^2 D\partial_r\rho(r,t)|_{r=R}$. \ 
The density profile $\rho(r,t)$ can be calculated easily as follows.
For the uniform initial condition, $\rho({\bf r}, t)$ is
only a function of the radial distance $r$ and it satisfies the diffusion equation
$\partial_t \rho= D\left[\partial^2_r \rho + (2/r)\partial_r\rho \right]$ for $r\ge R$
with the initial condition $\rho(r,0)=\rho_0$ and the boundary conditions
$\rho(R,t)=0$ and $\rho(\infty,t)=\rho_0$. \ 
One can reduce this problem
to a $1$D diffusion problem with $F(r,t) = r \rho(r,t)$ satisfying
$\partial_t F(r,t)= D \partial^2_r F(r,t)$, and
$F(r,0) = \rho_0 r$ and $F(r=R,t) = 0$,
whose solution can be easily found via the method of images. Dividing $F(r,t)$ by $r$
one gets the density profile 
\begin{equation}
\rho(r,t) = \rho_0 \left[1-\frac R {r} {\rm erfc} \frac{r - R}{\sqrt{4 D t}}\right]. 
\label{image}
\end{equation}
Consequently the flux is $\Phi(t)= 4\pi R D \rho_0[1+ R/\sqrt{\pi D t}]$ which tends
to a constant $\Phi(\infty)= 4\pi RD \rho_0$ at long times.
Comparing with Eq.\ (\ref{survival}) one sees that the survival probability
$S(r_0,t)$ has the same expression as the density profile $\rho(r,t)/\rho_0$
in the flux problem provided one replaces $r$ by $r_0$ in the latter problem.

To understand the origin of this connection between the two problems it is useful
to discuss both of them simultaneously in terms of a single Green's function.  
The Green's function $G({\bf r}, {\bf r_0}, t)$
for finding the particle at position ${\bf r}$ at time $t$, starting
at ${\bf r}_0$ outside the sphere at $t=0$, satisfies 
\begin{equation}
\frac{\partial G({\bf r}, {\bf r_0}, t)}{\partial t} =  D \nabla^2 G({\bf r}, {\bf r_0}, t) \ ,
\label{diffeq}
\end{equation}
subject to the initial condition $G({\bf r}, {\bf r_0}, 0) = \bf \delta(\bf r - \bf r_0)$,
and the absorbing boundary condition $G({\bf r}, {\bf r_0}, t) = 0$ for $|{\bf r}| = R$. \    
In the flux problem, given an arbitrary initial density $\rho({\bf r}_0, 0)$, 
the density at time $t$ can be found from
\begin{equation}
\rho({\bf r}, t)= \int G({\bf r}, {\bf r}_0, t) \rho({\bf r}_0, 0) d{\bf r}_0 \  
\label{green}
\end{equation}
where the integration is over $r_0>R$. \ 
In the capture problem, the survival probability 
$S({\bf r}_0,t)$ up to time $t$, starting at $ {\bf r}_0$, can also be written in terms of the
same Green's function,
\begin{equation}
S({\bf r}_0, t) = \int G({\bf r}, {\bf r}_0, t) d{\bf r}
\label{eqr0}
\end{equation}
where the integration is over $r\ge R$. \  For uniform and isotropic initial density $\rho_0$
in the flux problem one gets from Eq.\ (\ref{green})
\begin{equation}
\rho({\bf r}, t)= \rho_0 \int G({\bf r}, {\bf r}_0, t) d{\bf r}_0.
\label{eqr}
\end{equation}
Since the Green's function $ G({\bf r}, {\bf r}_0, t)$ is symmetric in ${\bf r}$ and ${\bf r}_0$
for unbiased Brownian motion, it follows by comparing Eqs. (\ref{eqr0}) and (\ref{eqr}) the
equality
\begin{equation}
S({\bf r}_0, t) =\rho({\bf r}_0, t)/\rho_0
\label{seqrho}
\end{equation}
This leads
to the quite general result that the probability $S({\bf r}_0, t)$ a particle at position
${\bf r_0}$ survives
up to time $t$  can be found directly by solving the density profile in
the corresponding
flux problem starting from an initial uniform density.
Eq.\ (\ref{seqrho}) just says that the net probability a
walker starting at point $r_0$ survives up to time $t$ is the same as the
sum of the probabilities that particles starting at every point in space reach
$r_0$ in time $t$ without being adsorbed. We note that the same general argument
goes through even for discrete-time random walks, leading to the relation
$S_n(r_0)= \rho_n(r_0)/\rho_0$, where $\rho_n(r_0)$ is the density profile
at step $n$ of the corresponding discrete-time flux problem.  This equivalence
is also discussed in Refs.\ \cite{Ziff91,Redner01}.

Thus we can use many of the discrete-time flux-problem results 
from Refs.\ \cite{Ziff91,MajumdarComtetZiff06,ZiffMajumdarComtet07}
to study the present capture problem.  Besides mapping those
results onto the present problem, we derive several new results, 
details of which will be given in \cite{MajumdarComtetZiff08}.
We also reformulate the mathematical expressions
in order to scale out the root-mean square (r.m.s.) jump length $\sigma$  from the integral expressions and 
leave them dimensionless~\cite{comment1}. 


We consider that a particle at position ${\bf r}'$ (with $r'>R$) jumps in one time step
(of duration $\tau$)
to a new position $\bf r$ with a jump distance $|{\bf r}-{\bf r}'|$ that is drawn
independently from an isotropic distribution $4 \pi r^2 W(|\bf r-\bf r'|)$, bounded above
by $2R$, and 
normalized as $4\pi \int_0^{2R} W(r) r^2 dr=1$, as illustrated in Fig.\ \ref{walkfig}.
Because of radial symmetry, this problem
can be formulated as a one-dimensional
problem, exactly analogous to the case of radial diffusion discussed above,
with effective density $\tilde \rho_n(r) = r \rho_n(r)$ and an effective jump probability
given by the symmetric, non-negative function
\begin{equation}
f(x)= 2\pi \int_{|x|}^{2R} W(r)\, r \, dr \ .
\label{kernel1}
\end{equation}
which is also normalized to unity, $\int_{-\infty}^{\infty} f(x) dx=1$. \    
The r.m.s. jump length of  $f$ is  $1/\sqrt{3}$ times the jump length for $W$: 
$\sigma^2= \int_{0}^{\infty} W(r)\, r^2\,  4\pi r^2 d{r}  
= 3 \int_{-\infty}^{\infty} f(x)\, x^2\, dx$. \   The quantity $\tilde \rho_n(r)$
satisfies 
\begin{equation}
\tilde \rho_{n+1}(r) = \int_{|r-\ell_m|}^{r+\ell_m} f(|r - r'|) \tilde \rho_n(r') dr'
\label{oned}
\end{equation}
 where $\ell_m$ is the maximum of the jump distance, and the subscript
 $n$ represents the time step.
 
 We rescale the jump distribution by $\sigma$ 
and thus define a new function $g(y)$ by $f(x) = (1/ \sigma) g(x/\sigma)$
so that $g(y)$ is normalized to unity and has a second moment of $1/3$. \ 
Two important quantities related to $g(y)$ appear:
\begin{equation}
 c = - \frac{1}{\pi}\, \int_0^{\infty}  \ln \left[\frac{1-\hat
g(k)}{k^2/6}\right]\, \frac{dk}{k^2}\,
\label{stat5}
\end{equation}
 and 
\begin{equation}
 b = -\frac{1}{\pi \sqrt{6}} \int_0^\infty \ln [1 - \hat g(k)]\,dk 
\end{equation}
where $\hat g(k) = \int_{-\infty}^\infty g(y) e^{i k y} dy$. \ 

\begin{figure}
\includegraphics[width=0.9\hsize ]{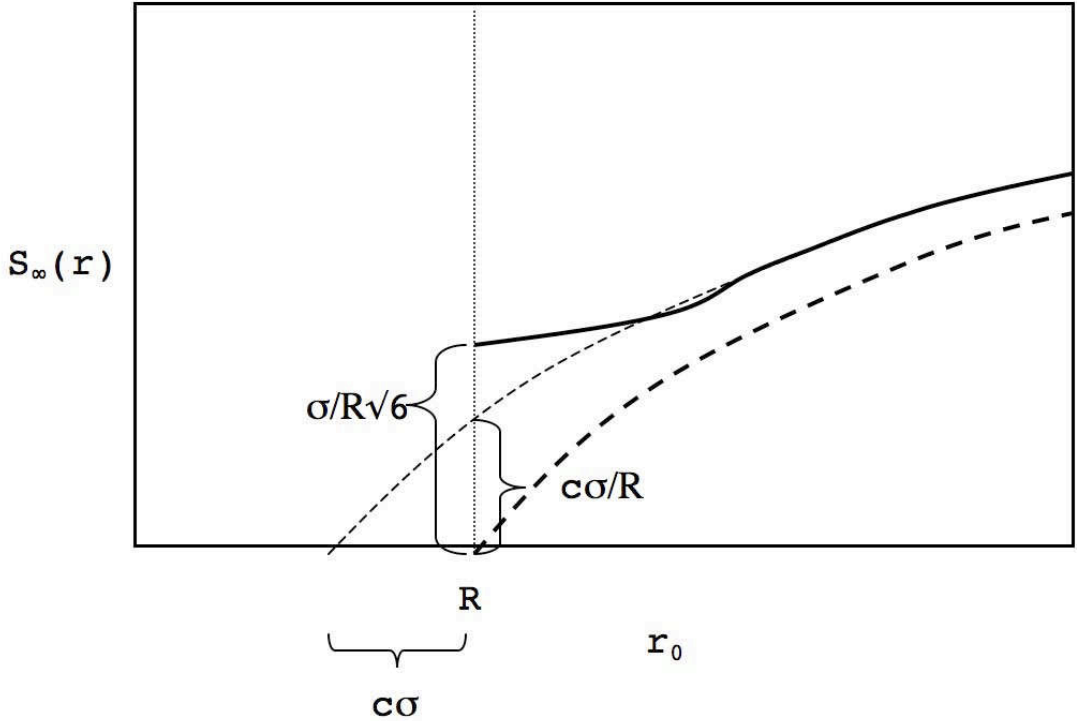}
\caption{
Behavior of the net (infinite-time) survival probability $S_\infty(r_0)$
as a function of the distance $r_0$ from the center of the adsorbing sphere
(solid line).
Also shown is the continuous-time prediction (heavy dashed line), 
and extrapolation of asymptotic behavior to $S_\infty = 0$ (light dashed line).
The extrapolation length is $ c \sigma$,
the actual value of $S_{\infty}(R)$ is $\sigma/(R \sqrt{6})$, while its extrapolated value
is $c \sigma/R$.}
\label{fig2}
\end{figure}

In \cite{ZiffMajumdarComtet07}, the general solution 
to Eq.\ (\ref{oned}) is given, in terms of a 
double Laplace transform of $S_n(r) = \rho_n(r)/\rho_0$. \ 
Here we analyze that result
explicitly in two important limiting cases:

(i) For $r_0 \gg R$ and for a large number of time-steps $n \gg 1$, the discrete-time
survival probability $S_n(r_0)$ behaves as
\begin{equation}
S_n(r_0) = 1 - \frac {R'} {r_0} {\rm erfc} \frac{r_0 - R'}{\sqrt{2 \sigma^2 n / 3}}  + {\cal O}(n^{-3/2})
\label{survival2}
\end{equation}
where $R' = R -  c \sigma$. \   This is valid for $r_0 \ll n \sigma$, in which case $r_0$ is
well within the maximum distance a particle can travel in $n$ bounded steps.
We can write the above result explicitly
in terms of time $t = n \tau$, and introduce the effective diffusion coefficient
\begin{equation}
D = \sigma^2/(6 \tau)
\end{equation}
which implies that the factor $2 \sigma^2 n / 3$ in (\ref{survival2}) equals $4 D t$. \ 
Then Eq.\ (\ref{survival2}) becomes identical to 
Eq.\ (\ref{survival}), except that the effective radius of the capture
sphere is reduced, as shown in Fig.\ \ref{fig2}.
The ultimate survival probability for $r_0\gg R$ is simply $S_\infty(r_0) = 1 - R'/r_0$
as given in Eq.\ (\ref{univ})



Note that if we take the Brownian limit of $\sigma \to 0$ and
$\tau \to 0$ with $D$ = const., then 
Eq.\ (\ref{survival2})  becomes Eq.\ (\ref{escape}) exactly.

(ii) For $n = \infty$, the Laplace transform of the steady-state solution, written in terms of $\hat g(k)$, simplifies to
\begin{eqnarray}
&& \int_0^\infty  F_\infty(z) e^{- \lambda z} d z \nonumber \\
 && = \frac{1}{\lambda \sqrt{6} } \exp \left[ -\frac{\lambda}{\pi} \int_0^\infty   \frac{\ln [1 - \hat g(k)]}{\lambda^2 + k^2}
 dk \right] 
 \label{laplace}
 \end{eqnarray}
where $F_\infty(z) = r_0 S_\infty(r_0)/\sigma$ and $z = (r_0 - R)/\sigma$. \ 
For $z \ll 1$,
Eq.\ (\ref{laplace}) implies $F_\infty(z) = 1/\sqrt{6} + b z \ldots$, which yields 
\begin{equation}
S_\infty(r_0) = \frac{\sigma}{r_0 \sqrt{6}}  +    \frac{ b (r_0 - R)}{r_0}+ \ldots
\label{asymp}
\end{equation}
for $r_0 - R \ll \sigma$. \ 
When $r_0 = R$, this gives Eq.\ (\ref{univ}).
If we set $\sigma = R$ (the mean jump length equal to the radius), the escape probability from the surface is 
$1/\sqrt{6} \approx 0.408248$. \   \  This compares with the  escape probability
$\approx 0.659463$ for a cubic lattice that follows from Eq.\ (\ref{cubic}).

It is interesting to note that as $\sigma \to 0$, the first term in Eq.\ (\ref{asymp}) vanishes,
but the second term does not.  Thus, no matter how small $\sigma/R$ is, in
the small region near the surface $R < r_0 \ll  R + \sigma$, the slope of the curve
of $S_\infty(r_0)$ vs.\ $r_0$ has the value $ b/R$ rather than the value
$1/R$ corresponding to the diffusion-equation solution.


Because the asymptotic flux $\Phi(t\to \infty)$ has a universal value $4 \pi D R \rho_0$ in the Brownian
limit, one might wonder whether the universality of $ S_\infty(R)$  given by 
Eq.\ (\ref{univ}) (now viewed in the context of the flux problem)
is related to that universal flux value. 
The flux $\Phi_n$ as $n\to \infty$ is found from \cite{MajumdarComtetZiff08}:
\begin{equation}
\Phi_{\infty} = \frac{4\pi \rho_0}{\tau} \int_R^{R + \ell_m} dr' r' S_\infty(r')\int_{r'-\ell_m}^R dr\, r 
f(r - r') 
\label{flux}
\end{equation}
and depends upon the
behavior of $S_\infty(r)$ within a distance $\ell_m$ of the sphere.   For $\Phi_{\infty}$ to give
$4 \pi D R \rho_0$ to leading order in $\sigma/R$, it follows that we must have
\begin{equation}
 \int_0^{\frac{\ell_m}{\sigma}} dz'  F_\infty(z')  \int_{z'-\frac{\ell_m}{\sigma}}^0 dz\,  g(z - z') = \frac{1}{6}
\label{constraint}
\end{equation}
But all orders in the expansion of $F_\infty(z') = 1/\sqrt{6} + b z' ...$
contribute to the integral in Eq.\ (\ref{constraint}), and so it follows that 
the universal leading term in Eq.\ (\ref{asymp}) is {\it not} simply a consequence
of the universality of the flux, making that term all the more intriguing.



The values of $ c$ and $ b$ can be calculated for various jump distributions.  The most natural
one from the point of view of computer simulation is the Pearson flight $W(r) = \delta(r - \ell)/(4 \pi \ell^2)$,
which leads to $f(x) = 1/(2 \ell)$ for $x \in (-\ell, \ell)$ or $g(y) = 1/2$ for 
$y \in (-1,1)$ (and zero otherwise), and $\hat g(k) = \sin k /  k $. \   Then
\begin{equation}
c = - \frac{1}{\pi}\, \int_0^{\infty}  \ln \left[\frac{1-(\sin \xi) / 
\xi}{\xi^2/6}\right]\, \frac{d\xi}{\xi^2}
\approx 0.2979521903  \label{integral} \end{equation}
and
\begin{equation}
b  =  -\frac{1}{\pi \sqrt{6}} \int_0^\infty \ln \left[1 - \frac{\sin \xi }{ \xi 
}\right]\,d\xi \approx 0.6538250956 \label{bfixed}
\end{equation}

The number $c = 0.29795\ldots$ has had a fairly long history.
It first appeared in \cite{Ziff91} in the context of the flux problem,
where its value was determined through a  
numerical iteration.  It later appeared independently
\cite{CoffmanFlajoletFlatoHofri98} in a study of the 
asymptotics of a sum of random variables with a uniform distribution,
and was evaluated by a slowly converging double
summation.  The integral form (\ref{integral}) was given in \cite{ComtetMajumdar05} in the context of 
the $1$D random-walk problem.  Finally, relation between
 the $3$D flux and $1$D random-walk problems was shown in \cite{MajumdarComtetZiff06}.

For the Pearson flight, we can find $F_\infty(z)$ to higher order numerically $F_\infty(z) = 1/\sqrt{6} + b z + 
0.2362658938 z^2 + 0.014221827913 z^3 \ldots$, and in this case
we confirm that  Eq.\ (\ref{constraint})  is indeed
satisfied to high precision (when more of these terms are included).
For the Pearson flight, the second integral in Eq.\ (\ref{constraint}) gives $(1-z')/2$. \   
Also, if we define $ I_k = \int_0^1 (1 - z)^k F_\infty(z) dz $, then we 
find  $I_0 = 2 / \sqrt{6}$, $I_1 = 1/3$, and $I_2 = 2 c/3$. \   We
also find $F_\infty(1) = 2b$. \ 

As a second example, we consider that the jump is uniform 
within a sphere of radius $\ell$, so that
$W(r) = 3/(4 \pi \ell^3)$ for $|r| < \ell$ and zero otherwise, implying $f(x) = (3/4)(\ell^2-x^2)/\ell^3$,
$g(y) = (3/4) \sqrt{3/5} (1 - 3 y^2/5)$ ($|y| < \sqrt{5/3}$), 
$\tilde g(k) = 9\sqrt{15} \sin(k\sqrt{5/3})/(25k^3) - 9 \cos(k\sqrt{5/3})/(5 k^2)$, 
$b = 0.682012\ldots$ and  $c = 0.310901\ldots$. \   Notice that the value of $c$ is just a little larger than that of the Pearson walk.  

\section{Further discussion and conclusions}
In deriving these results, we assumed that adsorption occurs only if the final position of the particle
falls within the sphere -- thus trajectories that pass through or graze the sphere are assumed not to adsorb.  
Including these events lead to  corrections of higher order in $\sigma/R$ \cite{Ziff91}.   Physically, of course, such trajectories should be adsorbed, so in that case our results are only accurate for $\sigma/R \ll 1$. \   However, for computer simulation it is easiest and therefore common to only check for adsorption based upon the final position, in which case our results are exact as long as $W(r) = 0$ for $r > 2R$. \  
	
The assumption $W(r) = 0$ for $r > 2R$ is necessary so that the transformation from the  $3$D to the $1$D problem described by (\ref{oned}) is exact.  Once on the $1$D level, however, the mathematics of the calculation holds for
any length jump distribution, and here we consider two infinite-range models.  If $\sigma \ll R$, then
the probability of jumping beyond $2R$ is very low and these results should give good approximations to the
true $3$D behavior.


In the first of these models, we consider the exponential distribution
$g(y) = \sqrt{3/2} \,  e^{-\sqrt{6} |y|}$, implying
$\hat g(k) = 6/(k^2 + 6)$, and find $c = 1/\sqrt{6} = 0.408248...$ and $b = 1$. \ 
Thus, $R' = R - \sigma/\sqrt{6}$, and by Eq.\ (\ref{asymp}),
$S_\infty(r_0) = \sigma/(r_0 \sqrt{6}) + (r_0 - R)/r_0
=  (r_0 - R +  c \sigma)/r_0$,
which in this case is the solution for all $r_0$, not just for $r_0 - R \ll \sigma$. \ 

Secondly, we consider a distribution that is Gaussian
in $W$ and therefore also in $g$. \ 
Then,  $g(y) = \sqrt{3/(2\pi)} \exp(-3y^2/2)$, $\hat g(k) = \exp(-k^2/6)$, and
$ c = -\zeta(1/2)/\sqrt{6 \pi} = 0.336363 \ldots $ 
 (similar to what was found in \cite{ComtetMajumdar05}
in another context) and $ b = \zeta(3/2)/(2 \sqrt{\pi}) = 0.736937\ldots$. \  
Interestingly, $\zeta(1/2)$ also appears in the problem of adsorption of a particle
(in $1$D) diffusing by the Ornstein-Uhlenbeck process\cite{MarshallWatson85,DoeringHaganLevermore87,HaganDoeringLevermore89},
where it is also related to the boundary extrapolation length, as well as in 
in the context of the maximum of a Rouse polymer chain~\cite{ComtetMajumdar05} or Gaussian random 
walks~\cite{JL}.  

Thus, we have shown that the discreteness of the time always affects the ultimate capture
probability of a particle undergoing a random walk.  Even if the walk length were drawn
from a proper Gaussian distribution appropriate for Brownian motion representing
the time interval $\tau$, the capture probability would not be the same as the solution for
the diffusion equation, because of the ability of the discrete walk to jump away
from the surface.   Different distributions
change the constants $ c$ and $ b$, but the general
behavior remains the same.

This work implies that, when doing computer simulations involving capture on a sphere, 
in order to get a 
capture probability that is say 99\% correct,
one should make $\sigma$ less than $(0.01/ c) R$,
or in the case of  the Pearson flight, make  $\ell$
less than $0.03356 R$ by Eq.\ (\ref{integral}).
 Indeed, most simulations use
much larger jump lengths and therefore significantly underestimate the
capture probability.  An improvement in accuracy can be achieved by varying 
the step size as the adsorbing boundary is approached \cite{MajumdarComtetZiff08}.

\section{Acknowledgments}
   Support of the National Science 
Foundation under Grant No. DMS-0553487, and from the Universit\'e Paris-Sud 11
for a visiting professorship, is gratefully acknowledged by RMZ.

\end{document}